\documentclass{iopart}
\usepackage{graphicx}
\usepackage{cite}

\begin{document}

\title{ The Future of Quark Matter at RHIC }

\author{Barbara V. Jacak and Michael P. McCumber}
\address{Department of Physics and Astronomy,
Stony Brook University,\\
Stony Brook, New York 11794-3800,
USA}
\ead{Barbara.Jacak@stonybrook.edu}


\bigskip

The matter discovered at RHIC has shown a number of very surprising
 properties. It is extremely dense and has an initial energy density
 well above that expected for the phase transition from ordinary
hadronic matter into quark gluon plasma.\cite{phxwp,starwp,phowp,brwp}
 The dense matter is very opaque to colored particles, such as quarks
 or gluons, traversing it. This gives rise to the striking jet
 quenching observed at RHIC.\cite{ppg003,starjetquench}. Measurements
 of the elliptic flow of different hadrons indicate that the plasma at
RHIC is a liquid with extremely low
 viscosity.\cite{phxwp,starwp,phowp,brwp} Furthermore, there is
 substantial evidence that hadron formation results from coalescence of
 constituent quarks from the collectively flowing medium.\cite{reco}

It is natural to ask how this hot, dense partonic liquid transports
 particles, energy, momentum and charge. This means we should determine
 its diffusion constant, thermal conductivity, viscosity and
 (color)electric conductivity. Extracting such quantities
 quantitatively requires significant advances in both fundamental
 theory and phenomenology. Of course, these must be accompanied by
 extensive, high precision measurements.
In this paper, we will discuss how increased luminosity at RHIC, along 
with upgraded detector capabilities, will probe how
 this new kind of plasma works.

Upgrading the luminosity at RHIC is compelled by two kinds of questions.
 There are basic questions which for which we don't yet have sufficient
 data. And there are entirely new questions, which arose from the
 initial discoveries at RHIC. Such new questions include:
\begin{itemize}
\item What is the mechanism for rapid thermalization?
\item How low is the viscosity of the liquid?
\item How does the plasma respond to deposited energy?
\item What is the color screening length?
\item Is the initial state a color glass condensate?
\end{itemize}
Early questions not yet answered by data are:
\begin{itemize}
\item What is the nature of phase transition? Where is the critical point?
\item What is the equation of state of hot QCD matter?
\item Do heavy quark bound states melt?
\item Can dilepton observables provide evidence for chiral symmetry restoration?
\end{itemize}

Lattice QCD indicates that the phase transition should occur above an
 energy density of 0.6 GeV/fm$^3$ and temperature of 150 MeV.\cite{lattice}
The ratio
of energy density to temperature, $\epsilon/T^4$ increases rapidly, and
 then levels off within the range 150 $\le T \le$ 300 MeV. The RHIC
 collider thus can probe exactly the region of interest. The addition
 of electron cooling of the heavy ion beams will counteract beam
blow-up due to Coulomb interactions in the dense beams of highly
charged particles.  Cooling will increase the luminosity for heavy
 ion collisions by an order of magnitude. This will allow, among other
 things, a scan of colliding beam energies to search for the critical
 point with good statistical precision. In addition, completion of the
 new Electron Beam Ion Source (EBIS) will offer the first ever
collisions of uranium beams. U+U collisions with
 high luminosity will allow probing systems with 30\% higher energy
 density, by selecting the collisions in which 
the non-spherical uranium nuclei are aligned end-on.

Both the STAR and PHENIX experiments have undertaken a number of
 detector upgrade projects. These will maximize the benefit from higher
 collider luminosity, but will also greatly increase the physics reach of
 the experiments even before the luminosity upgrade is complete. STAR
 is currently implementing a large barrel time-of-flight upgrade to
 increase particle identification capabilities, a forward meson
 spectrometer to allow measurement of neutral mesons at forward 
rapidity, and preparing a data acquisition and TPC readout upgrade to 
enhance STAR's data acquisition rate capability. In the near future, 
STAR plans to
 construct a tracking upgrade consisting of a heavy flavor tracker
 (HFT) pixel detector, barrel and forward silicon trackers, and a 
forward triple-GEM EEMC tracker. PHENIX is constructing a Hadron Blind
 Detector for background rejection under low mass dileptons, a Muon
 Piston Calorimeter at forward rapidity, and a barrel silicon vertex
 tracker to identify open heavy flavor decays. In the future, work will
 begin on a forward vertex tracker and compact calorimeter extending
 the rapidity reach of PHENIX for photons and neutral mesons to the
 region 1 $\le y \le$ 3. The combination of calorimetry and tracking
 will give PHENIX a large acceptance for $\gamma$-jet coincidences.

Better determination of the viscosity and equation of state of the
 plasma at RHIC will require improved measurements of radial, directed,
 and elliptic flow for different mass hadrons. The particle mass
 dependence reflects the equation of state. The flows of multi-strange
 hadrons will help to differentiate late stage dissipation from early
 viscosity. 
Heavy meson elliptic flow and diffusion through the plasma will pin 
down the thermalization time and viscosity, and show how these vary 
with beam energy. The detector upgrades described above will greatly
 improve the particle identification capabilities of both STAR and
 PHENIX to make this possible.
Of course, for these measurements to produce precision science requires that RHIC
scan in energy and system size (which will become possible with the
 higher luminosity of RHIC II), as well as careful comparison to
 viscous 3-dimensional hydrodynamics calculations. Significant progress
 in theory and phenomenology will be needed to make this possible.
 Furthermore, fundamental theory breakthroughs are needed to 
quantitatively understand
 the initial state in heavy ion collisions, pin down the thermalization
 mechanism, identify experimental observables of instabilities, and
 calculate pre-equilibrium dynamics of the matter.

In traditional plasma physics, it is common to use externally
generated probes and measure their transmission through the
plasma. At RHIC,
determining the response of our plasma to deposited energy requires
something equivalent to the plasma physicists' ``external" probes. 
Heavy ion collisions provide just such probes, autogenerated 
in the collision itself, arising in the initial hard scatterings
among partons in the nuclei.
The production rate and distribution are calculable with
perturbative QCD, and are experimentally well calibrated
in p+p collisions. These hard probes
 interact with the medium as they traverse it;
jets and heavy quarks are of particular interest.
 Experimentally, the goal is to measure the fate of the probe, along
 with associated plasma-generated particles. 

Transport in plasmas is
 driven primarily by collisions. Diffusion is a measure of the
 transport of particles by the plasma. The transport of energy by those
 particles is given by the thermal conductivity. Viscosity is a
 measure of the transport of momentum by particles, while electrical
 conductivity measures how effectively the particle collisions transport
 charge. Precise determination of transport properties will require
 substantial progress in theory, in particular development of
 techniques to handle transport in a medium where the coupling between
 particles is not weak. A number of ideas were discussed at this
 conference, ranging from next-to-leading order perturbative
 calculations of energy loss in an expanding medium, 
resummation techniques to allow the application of 
perturbation theory to
 strongly coupled regimes, scattering of particles from color fields,
 possibly coherent, and application of AdS/CFT correspondence to allow
 use of gravitational calculations in the infinitely strongly coupled
 limit. It is clear that we are only beginning to learn how to make the
 problem tractable.

Existing data have established the usefulness of heavy quarks and jets
 as good candidate plasma probes, but current energy reach and
 precision are far short of what is required. As these are rare
 processes with triggerable signatures, the RHIC II luminosity will
 provide an enormous improvement. There are three basic measurements
 required. One is spectroscopy of heavy quarkonia, to determine the
 screening length in the plasma and fate of heavy bound states. This
 will require measurement of $J/\psi, \psi\prime$ and $\Upsilon$ states.
 Another is to measure the correlations of hard particle pairs to
 study the mechanism by which fast quarks or gluons lose energy to the
 medium. High statistics measurements of hadron fragments with $p_T
 > 5$ GeV/c arising from 10-30 GeV jets are needed for substantial
 progress. The medium response to the lost energy is probed by two and
 three-particle correlations, with a hard trigger particle and medium-generated associated particles with 1 $\le p_T \le$ 4 GeV/c.

It is natural to wonder about the role of RHIC II, as hard probes are
 major goals of the heavy ion program at the LHC. At RHIC II,
the high
 luminosity, long running time, and flexibility of the accelerator
 complex provide an extremely compelling -- and competitive -- physics
 program with hard probes. The higher cross section at LHC energy is
 more than compensated by the larger integrated luminosity at RHIC II. 

J/$\psi$ suppression was predicted 20 years ago as a signature of
 deconfined matter~\cite{matsui_satz}.
Since then, charmonium production has been extensively studied.
At the CERN SPS, NA50 measured ``normal'' J/$\psi$ suppression in 
cold nuclear matter, {\it i.e.} in p+A collisions, and ``anomalous'' 
suppression in A+A collisions.\cite{na50} The observed decrease of the 
J/$\psi$ yield in central A+A collisions has been interpreted as
indicating deconfinement. However, this interpretation is complicated
by the fact that PHENIX observes a similar level of suppression
despite considerably higher energy density at RHIC.\cite{phx_jpsi}
It is conceivable
that the suppression signals sequential melting of
different charm quark bound states, along with feeding by decays
of heavier states. Indeed, such a result would be expected for a 
strongly coupled plasma in which the $c$ and $\overline{c}$ potential
is not fully screened. However, it may also be that the primordial 
J/$\psi$ is more suppressed at RHIC than at the SPS and the additional 
suppression
is canceled by final state coalescence of charm and anti-charm quarks.
Sorting this out unambiguously will require measuring multiple bound
states for several beam energies and systems. These measurements, with 
higher precision and significantly larger $p_T$ reach, should
 allow determination of the color screening length in the plasma by
comparing the fate of bound states with different radii. Significant
theory work is needed as well, if we are to fully understand the role
of apparently canceling processes. Calculations need to be done that
include a well-define initial state (for example, color glass condensate),
color screening, and bound state regeneration in an expanding
medium. The effect of strong coupling in the medium upon heavy quark
bound states needs quantitative exploration; this has begun using
lattice QCD, but dynamics are an important component.

\begin{center}
\begin{figure}
\includegraphics[width=\textwidth]{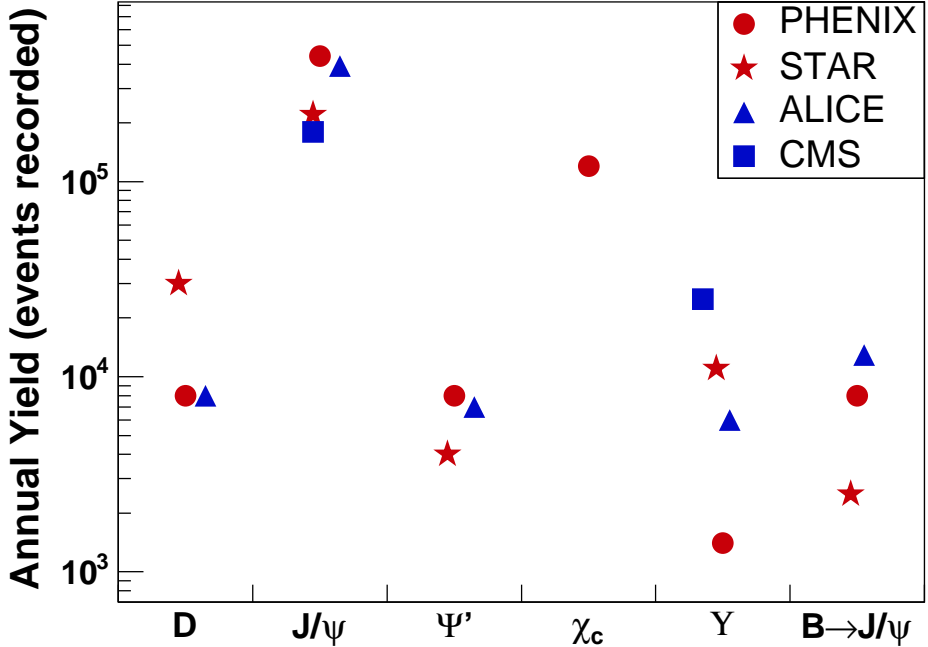}
\caption{ Expected annual yield for various mesons containing heavy
 quarks detected in minimum bias collisions of the heaviest ions. The
 PHENIX and STAR estimates (red symbols) are for Au+Au at RHIC II
 luminosity, in a run of 12 weeks duration. ALICE and CMS estimates
 (blue symbols) are for Pb+Pb collisions at LHC with design luminosity
 and 1 month of data taking.}

\label{fig1}
\end{figure}
\end{center}

Figure 1 shows the expected number of events with detected heavy flavor
 in one run for PHENIX and STAR at RHIC II and for the ALICE
 and CMS experiments at the LHC. It should be noted that the
RHIC II run length used is 12 weeks, which corresponds
to approximately one half of a year's worth of beam time. 
Good triggering capability is assumed
 for all four experiments. The rates are for minimum bias Au+Au and
 Pb+Pb collisions at RHIC II and LHC, respectively. It is clear that
 the number of heavy particles recorded will be quite comparable at the
 two facilities. This seems surprising, as the production cross
 sections are higher by a factor of 10-50 at the LHC, depending on the
 species. The similarity arises because the RHIC II luminosity will be
 a factor of 14 higher, and the running 3-4 times longer. 

It should be noted that charmonium states will be of limited utility as
 probes of color screening at the LHC. Because of the high initial
 temperature, there will be substantial thermal production of charm
 quarks, and the majority of the charm anti-charm bound states should
 be formed by final state coalescence. Consequently, charmonia at LHC
 will not provide detailed information about the early state of the
 medium at its hottest and densest. This role will be open to B quark
 states only. The planned $\chi_C$ measurement by PHENIX is very
 challenging due to large backgrounds. It is expected that the RHIC II
 sample will allow statistical separation of the signal from the
 background. At the LHC, the background will be larger still, most
 likely proving prohibitive for this measurement.

Turning now to jet probes of energy loss and transport in the hot and
dense medium, we compare the physics reach at RHIC II and LHC.
 Correlations of two or more high momentum particles arising from 
 fragmentation of back-to-back jets
probe the mechanism of energy loss. Such information can
 be gleaned from 10-30 GeV jets. Because of the lower temperature and
 smaller soft, or thermal, hadron backgrounds, this physics can be
 probed with 10 GeV jets at RHIC II. The backgrounds will be
larger and have a harder spectrum at the LHC, requiring jet
energies of $\approx$ 30 GeV in order to achieve signal
to background ratios comparable to those at RHIC. 
It should be noted, however, that the collective
boost imparted by the jet to co-moving plasma particles is
relatively modest. Consequently, it may be that the
plasma response information is most prominent in
relatively soft jets. 

\begin{center}
\begin{figure}
\includegraphics[width=\textwidth]{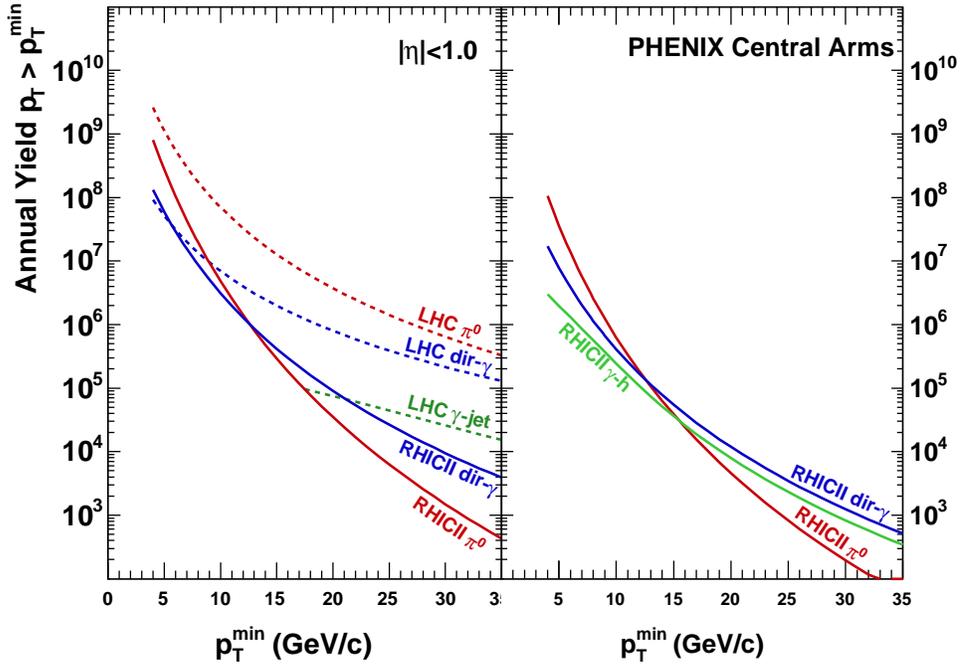}
\caption{ Annual recorded number of events of neutral pions, direct
 photons, and photon-jet coincidences at RHIC II and LHC, based upon
 NLO pQCD from W. Vogelsang and scaled to minimum bias Au+Au or Pb+Pb
 collisions (see text for details). The left panel shows yields into
 two units of rapidity centered at y=0 and full azimuth, while the
 right panel shows yields into the PHENIX central arms.}

\label{fig2}
\end{figure}
\end{center}

The correlation of one hard particle
 ($>$ 5 GeV/c $p_T$) with one or more softer particles ($p_T$  =
 1-4 GeV/c) from the opposing jet probes the response of the medium to
 the energy deposited by a jet traversing it. The ``golden channel" for
 jet probes of the hot, dense medium is direct photon-jet correlations.
 $direct \gamma$-hadron pairs arise from QCD Compton scattering; the direct 
photon tags the jet energy (i.e. the energy of the recoiling quark).
By measuring the hadrons from the jet, it should be possible to quantify 
the lost energy and its fate in the medium.

We estimate the expected annual rates using NLO perturbative QCD
 calculations from W. Vogelsang,\cite{vogel} which have been
 benchmarked by comparison to p+p collisions at RHIC. To predict rates
 in minimum bias heavy ion collisions at RHIC II and the LHC, the pQCD
 cross-sections of $\pi^0$ and direct photons are integrated above a
 given $p_T$ threshold, scaled up by $A^2$, and multiplied by the
 expected integrated luminosity. The resulting annual yields are
 corrected for the observed hadron suppression and direct photon 
non-suppression in Au+Au collisions at RHIC, and known or expected
 experiment up time.  In lieu of a measured suppression at the LHC we 
have used the RHIC value of 0.2 also for the LHC calculations. 
The plotted annual yields show the total number
 of recorded events with $p_T  > p_T^{min}$. Results for single
 direct $\gamma$ and $\pi^0$ can be seen in Figure 2 as solid lines for
 RHIC II and dashed lines for LHC. The left hand side shows the annual
 yield into rapidity -1 $ < y <$ 1, while the right hand
 side provides the yield into the PHENIX central arms for comparison.
 The larger rapidity range reflects the acceptance of STAR and ALICE,
 however PHENIX will have comparable acceptance once the upgrades are
 complete. 

Two striking results are immediately apparent. The first is that with 
the $\gamma/\pi^0$ ratio observed at RHIC, direct photons dominate over 
decay photons above $\approx$12 GeV/c $p_T$, greatly simplifying 
$\gamma$-jet measurements. The second is that the $p_T$ reach at RHIC 
II is quite large. Using 1000 counts as a useful minimum, we see that
 direct photons can be measured to $p_T > $ 30 GeV/c at RHIC II.

Yields of events with hadrons of $p_T \ge$ 4 GeV/c detected in
 coincidence with direct photons are also shown in Figure 2.  The
 coincidence probability of single hadrons opposing the direct photon
 or $\pi^0$  trigger in azimuth was estimated with the PYTHIA event
 generator to be essentially 1.0 for the left panel. The PHENIX event
 yields are based upon measured conditional yields of away-side hadrons,
 with PYTHIA used to estimate coincidence rates for higher
 $p_T$  trigger particles. The figure shows that RHIC II will measure
 direct photon-hadron coincidences for 35 GeV/c photons. Even the
 PHENIX central arms alone will access $direct \gamma$ - h for 30 GeV/c
 photons, though upgrades underway will increase the PHENIX acceptance
 nearer to that in the left panel. 

The LHC $\gamma$-jet curve shows expected yields for fully
 reconstructed away side jets. There will be a good sized sample of
 such events after one year of running. However, it should be noted
 that at LHC $\gamma/\pi^0$  remains significantly below 1.0, even at
 $p_T$  =35 GeV/c, unless the $\pi^0$  suppression at LHC were to be
 much larger than that observed at RHIC. This would be surprising since
 $R_{AA}\approx $ 0.2, observed at RHIC, is consistent with emission
 only from a thin shell around the surface of the medium. Consequently
 a significantly larger number of events will be required at the LHC,
 in order to enable statistical subtraction of decay photon-jet
 correlations. At RHIC, full jet reconstruction is not available for
 the relatively low energy jets measured thus far, but reconstruction
studies are currently underway.

\begin{center}
\begin{figure}
\includegraphics[width=\textwidth]{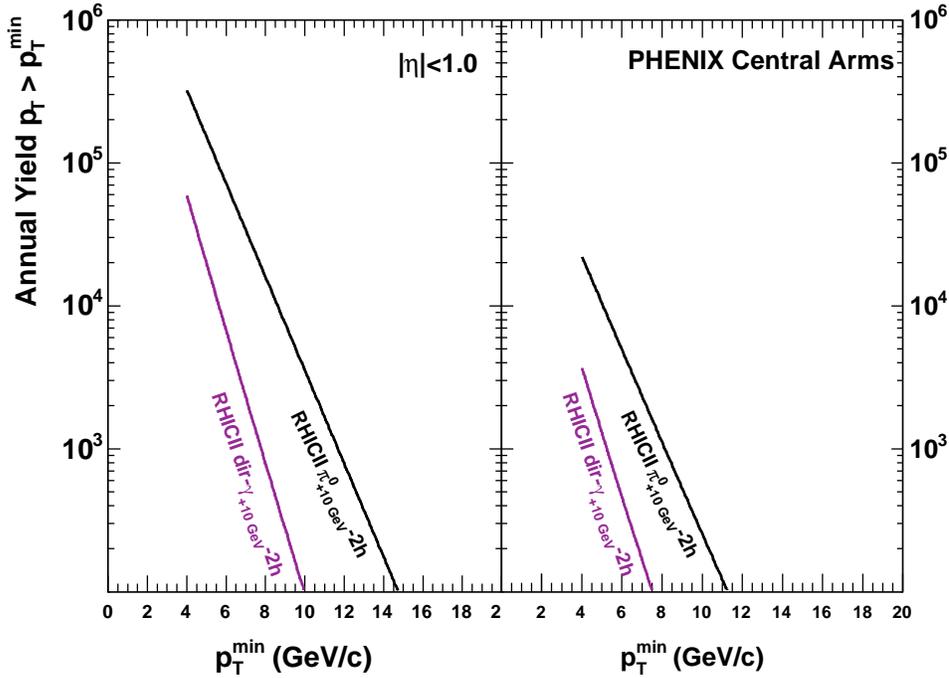}
\caption {Annual yield of events with two hadrons on the
opposing side of a $>$10 GeV/c
 direct photon and 10 GeV/c $\pi^0$. The event count is plotted as a
 function of hadron $p_T^{min}$.}

\label{fig3}
\end{figure}
\end{center}

In order to estimate the yield of events at RHIC II with more than a
 single hadron correlated with trigger jets and direct photons, we have
 performed PYTHIA simulations. Figure 3 shows annual yields of events
 with two hadrons on the away side of a $\ge$10 GeV/c direct photon and
 $\ge$10 GeV/c $\pi^0$. The number of events with two detected hadrons of 
 $p_T \ge  p_T^{min}$ is plotted as a function of hadron $p_T^{min}$.
 For such low energy jets, two hadrons above 4 GeV/c $p_T$ already
 carry most of the jet energy. The plot shows that a significant event
 sample will be collected with nearly fully reconstructed away-side
 jets at RHIC. For higher energy direct photons, the probability of 
finding two energetic hadrons should increase. Consequently, a significant number of direct photon-two
 hadron correlations should be expected also for the higher photon
 energies. The direct photon-2h yields fall more rapidly with hadron
$p_T$ than $\pi^0$-2h, as the $\pi^0$ carries only a fraction of the
jet energy. Photons arise from lower energy jets than a
$\pi^0$ at the same $p_T$, and therefore are less likely to have
two associated energetic hadrons.

In order to achieve these goals, an upgrade strategy for the RHIC
experiments and collider has been defined. Both STAR and PHENIX are
constructing short-term upgrades, which will begin taking data in 
the time frame of 2007-2009. Concurrent with these are steps to
increase the RHIC luminosity through improvements in vacuum quality
and increasing the number of bunches in the machine. The new
EBIS source should be commissioned in 2009. 
Larger scale upgrades for both experiments
will begin construction in 2007 and 2008, in order to be complete 
before 2011. The detector upgrades will be available for
data-taking prior to the start of the heavy ion program at the LHC.
Brookhaven envisions beginning construction work on the RHIC II luminosity
upgrade around 2009 or 2010, with comissioning three years later. 
The enabling technology for RHIC II is electron cooling of the ion 
beams to mitigate beam blow-up from Coulomb scattering. It is interesting
to note that the electron cooling that increases RHIC's luminosity
ten-fold also positions RHIC for use as an electron-ion collider.
This would entail
adding a 20 GeV electron accelerator, along with a new detector,
 to the complex.

\ack
We thank the organizers for the invitation to speak about the physics program 
planned for RHIC II. We would like to express our sincere appreciation to Werner
Vogelsang and Peter Jacobs for the calculations of hard scattering rates at
RHIC and LHC energies made available to us. We are grateful to Tony Frawley, 
Thomas Ullrich, Bolek Wyslouch and Brian Cole for providing expected measured 
yields of quarkonia at RHIC and LHC. Thanks are due to the STAR and PHENIX
Collaborations for sharing their RHIC II projections with us. We benefited
from many discussions with Axel Drees, Tom Ludlam, and Sam Aronson.

\end{document}